\documentclass[aps,pra,twocolumn,groupaddress,longbibliography]{revtex4-1}

\usepackage{amsmath}
\usepackage{amsfonts}
\usepackage{amssymb}
\usepackage{graphicx}
\usepackage{color}
\usepackage{accents}
\usepackage[colorlinks=true, citecolor=cyan]{hyperref}

\newcommand{\ket}[1]{\left|#1\right\rangle}
\newcommand{\bra}[1]{\left\langle #1\right|}

\makeatletter 
\newsavebox{\@brx}
\newcommand{\llangle}[1][]{\savebox{\@brx}{\(\m@th{#1\langle}\)}%
  \mathopen{\copy\@brx\kern-0.5\wd\@brx\usebox{\@brx}}}
\newcommand{\rrangle}[1][]{\savebox{\@brx}{\(\m@th{#1\rangle}\)}%
  \mathclose{\copy\@brx\kern-0.5\wd\@brx\usebox{\@brx}}}
\makeatother

\newlength{\dhatheight} 

\newcommand{\qed}{\nobreak \ifvmode \relax \else
      \ifdim\lastskip<1.5em \hskip-\lastskip
      \hskip1.5em plus0em minus0.5em \fi \nobreak
      \vrule height0.75em width0.5em depth0.25em\fi}

\begin{document}

\title{Enhancing the effect of quantum many-body scars on dynamics by minimising the effective dimension}
\author{Shane Dooley}
\email[]{dooleysh@gmail.com}
\affiliation{Dublin Institute for Advanced Studies, School of Theoretical Physics, 10 Burlington Rd, Dublin, Ireland}
\author{Graham Kells}
\affiliation{Dublin Institute for Advanced Studies, School of Theoretical Physics, 10 Burlington Rd, Dublin, Ireland}
\date{\today}

\begin{abstract}
  Quantum many-body scarring is believed to be the mechanism behind long-lived coherent oscillations in interacting Rydberg atom chains. These persistent oscillations are due to the large overlap of the many-body scars with certain initial states. We show that the ``effective dimension'' is a useful measure for identifying non-thermalising initial states in many-body scarred systems. By minimising the effective dimension we find physically reasonable initial states of the Rydberg chain that lead to more pronounced and longer lived oscillations, accentuating the effect of the many-body scars on the dynamics.
\end{abstract}


\maketitle

\section{Introduction}

Due to advances in experimental techniques with ultracold atoms \cite{Blo-08, Ber-17}, trapped ions \cite{Bri-12}, nitrogen-vacany centres \cite{Zhu-11, Cho-17b}, and other platforms \cite{Xia-13, Kur-15}, it is now possible to probe the quantum coherent dynamics of interacting many-body systems. This opens the door to the exploration of new frontiers in condensed matter physics \cite{Geo-14}, but may also have a technological impact since it can help us to understand equilibration and thermalisation in quantum systems, which in turn may help us to develop strategies to protect quantum coherence in many-body systems.

Quantum equilibration and thermalisation are often studied by means of a quantum quench, whereby a closed system is initially prepared in the ground state $\ket{\psi(0)}$ of some pre-quench Hamiltonian $\hat{\mathcal{H}}^\text{pre} \equiv \hat{\mathcal{H}} (t < 0)$. After a sudden change of the Hamiltonian parameters at $t=0$, dynamics is then generated by the post-quench Hamiltonian $\hat{\mathcal{H}}^\text{post} \equiv \hat{\mathcal{H}}(t>0)$. The system equilibrates if all few-body observables settle to their equilibrium values, and stay close to these values for most times $t>0$. The system is said to have thermalised if all of these observable equilibrium values are approximately equal to their expectation values in the Gibbs state \cite{Mor-18}.



Of course, it is possible to slow down or avoid equilibration by starting from a non-equilibrium initial state $\ket{\psi(0)}$ that is a superposition of some subset of the post-quench Hamiltonian eigenstates with approximately resonant energy gaps. However, such an initial state is typically difficult, or impossible to create experimentally for a many-body system, particularly for a non-integrable $\hat{\mathcal{H}}^\text{post}$, as the corresponding pre-quench Hamiltonian may require highly non-local terms, or fine-tuning of a large number of Hamiltonian parameters \cite{Rei-08}.


Despite this, a recent experiment on a chain of cold Rydberg atoms found unexpected long-lived coherent oscillations for a certain initial state \cite{Ber-17}. The failure to equilibrate over experimentally long timescales for this initial state was argued to be due to its large overlap with a band of special eigenstates of the post-quench Hamiltonian \cite{Tur-18a, Tur-18b}. These special states violate the eigenstate thermalisation hypothesis (ETH) \cite{DAl-16}, and were dubbed \emph{quantum many-body scars}, due to similarities with quantum scars in single particle billiard systems \cite{Hel-84}. A large amount of recent work has uncovered various properties of scars in the Rydberg chain \cite{Khe-19, Ho-19, Cho-19, Lin-19, Bul-20} and in other models \cite{Mou-18a, Mou-18b, Bul-19, Ok-19, Sch-19, Iad-20, Mar-20, Shi-20}. 


In this paper, we investigate whether the failure to equilibrate can be made more extreme with initial states $\ket{\psi(0)}$ that are a superposition of a \emph{smaller} number of scar states of $\hat{\mathcal{H}}^\text{post}$. We find that minimisation of the \emph{effective dimension} is a useful way of targeting such states. For the Rydberg chain and a pre-quench Hamiltonian with next-nearest neighbour terms, we find that the number of participating scar states can be significantly decreased. The resulting dynamics are qualitatively different depending on whether $L/2$ is odd or even, where $L$ is the length of the chain. However, in either case the Loschmidt echoes are enhanced, and longer lived than for the initial states considered in Refs. \cite{Ber-17, Tur-18a}.

The layout of our paper is as follows. In Section \ref{sec:background} we briefly review the phenomenon of quantum many-body scarring in the Rydberg atom chain and introduce our figure-of-merit, the effective dimension. In Section \ref{sec:pre_quench_H} we introduce our pre-quench Hamiltonian, which involves next-nearest neighbour interaction between Rydberg atoms, and we discuss some of its ground state properties. Our main results are presented in Section \ref{sec:results}, where we show that this modification can lead to a significantly reduced effective dimension and a slowdown in equilibration.

\section{Quantum many-body scars in the Rydberg atom chain}\label{sec:background}

\begin{figure*}
  \includegraphics[width=2\columnwidth]{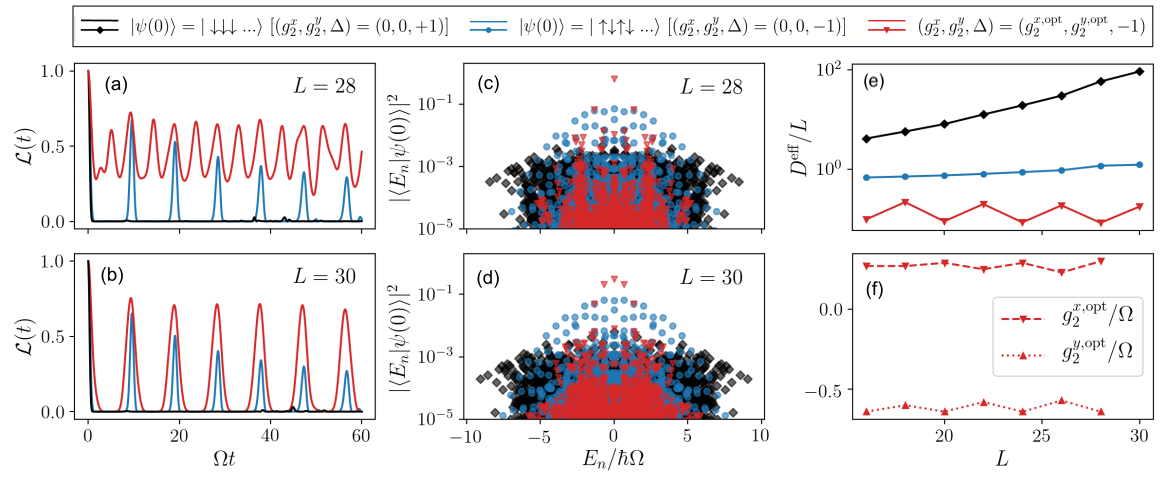}
  \caption{We compare evolution by the post-quench Hamiltonian $\hat{\mathcal{H}}_0^\text{post}$ for different initial states, specified by the pre-quench Hamiltonian parameters $(g_2^x, g_2^y, \Delta)$ [see Eqs. \ref{eq:H_0_pre}, \ref{eq:H_pre}, \ref{eq:delta_H_pre_2}]. (a, b) The initial state that minimises $D^\text{eff}$ (red lines) gives enhanced, broadened, and longer lived Loschmidt echoes, with qualitatively different dynamics if $L/2$ is even (a) or odd (b). (c, d) The optimal initial state gives an enhanced overlap with scar states in the middle of the spectrum, as well as a narrower energy distribution. If $L/2$ is even a single eigenstate dominates (c), while if $L/2$ is odd there are three eigenstates with significant overlap (d). (e) The optimal initial state gives a significantly reduced effective dimension $D^\text{eff}$, and a sub-linear increase with system size $L$. (f) The optimal pre-quench Hamiltonian parameters vary only slightly with $L$.}
  \label{fig:1}
\end{figure*}

An effective model for a chain of $L$ cold Rydberg atoms in the regime of nearest neigbour Rydberg blockade is given by the post-quench Hamiltonian \cite{Les-11}: \begin{equation} \hat{\mathcal{H}}_0^\text{post} = \hat{P} \hat{h}_0^\text{post} \hat{P} , \quad \hat{h}_0^\text{post} = \frac{\hbar\Omega}{2} \sum_{j=1}^L \hat{\sigma}_j^x , \label{eq:H_0_post} \end{equation} where $\hat{\sigma}_j^x = \ket{\uparrow_j}\bra{\downarrow_j} + \ket{\downarrow_j}\bra{\uparrow_j}$ acts on the two-level atom at site $j$ of the chain. The projector \begin{equation} \hat{P} = \prod_{i=1}^{L} \left( \hat{\mathbb{I}} - \ket{\uparrow_i\uparrow_{i+1}}\bra{\uparrow_i\uparrow_{i+1}} \right) , \label{eq:P} \end{equation} implements the blockade by excluding states $\ket{\hdots \uparrow\uparrow \hdots}$ with two consecutive $\uparrow$-states. Such states are annihilated by $\hat{P}$, and are therefore trivial zero-energy eigenstates of the Hamiltonian $\hat{\mathcal{H}}_0^\text{post}$ and may be neglected \footnote{As well as these \emph{trivial} zero-energy eigenstates, for which $\hat{P}|E \bigr> = 0$, there are also an exponential (in $L$) number of \emph{non-trivial} zero-energy eigenstates, for which $\hat{P}|E \bigr> \neq 0$ \cite{Tur-18a}. To avoid numerical instabilities associated with this degeneracy at zero-energy, our simulations also include a very small degeneracy-breaking perturbation to the post-quench Hamiltonian, of the form $\delta\hat{h}^\text{post} = \epsilon\sum_j\hat{\sigma}_j^z$, where $\epsilon = 10^{-9}\Omega$}. We note that a lower-case letter is used to distinguish the projected Hamiltonian $\hat{\mathcal{H}}_0^\text{post}$ from the Hamiltonian $\hat{h}_0^\text{post}$ without nearest neighbour blockade.

Assuming periodic boundary conditions, the Hamiltonian $\hat{\mathcal{H}}_0^\text{post}$, is invariant under the translation of site index $j \to j+1$, implying that the momentum $k = 2\pi n / L$, $n \in \{-L/2 + 1, -L/2 + 2, \hdots , L/2\}$ is a conserved quantum number. Reflection around the mid-point of the chain $j \to L - j + 1$ is also a symmetry of the Hamiltonian, implying the conservation of spatial parity $p=\pm 1$. Moreover, the post-quench Hamiltonian obeys a particle-hole symmetry $\{ \hat{\mathcal{H}}_0^\text{post}, \hat{\Pi} \} = 0$, where $\hat{\Pi} \equiv \bigotimes_{j=1}^L \hat{\sigma}_j^z$ is the excitation number parity operator. This implies that for any eigenstate $\ket{E}$ of $\hat{\mathcal{H}}_0^\text{post}$ with eigenvalue $E$ there is also an eigenstate $\hat{\Pi}\ket{E}$ with the eigenvalue $-E$.

It was observed in Ref. \cite{Ber-17} that for the initial state $\ket{\psi(0)} = \ket{\downarrow\downarrow\downarrow\hdots}$ the system rapidly thermalises. However, for the initial N\'{e}el state $\ket{\psi(0)} = \ket{\mathbb{Z}_2} \equiv \ket{\uparrow\downarrow\uparrow\downarrow\hdots}$ (or, alternatively, for $\ket{\psi(0)} = \ket{\mathbb{Z}'_2} \equiv \ket{\downarrow\uparrow\downarrow\uparrow\hdots}$) the dynamics show persistant oscillations of local observables. The blue lines in Fig. \ref{fig:1}(a, b) show the revivals of the Loschmidt probability $\mathcal{L}(t) \equiv |\langle\psi(0)|\exp(-it\hat{\mathcal{H}}_0^\text{post}/\hbar)|\psi(0)\rangle|^2$, calculated numerically for chain lengths $L \in \{28, 30\}$ and for $\ket{\psi(0)} = \ket{\mathbb{Z}_2}$. By comparison, for $\ket{\psi(0)} = \ket{\downarrow\downarrow\downarrow\hdots}$ the Loschmidt probability decays rapidly and does not revive within the time of our numerical calculation (the black lines). [The red lines in Fig. \ref{fig:1}(a, b) show the enhanced Loschmidt echoes for a modified initial state, which will be described in more detail in the following sections.]

Within a momentum/parity symmetry sector, the eigenvalue level statistics of $\hat{\mathcal{H}}_0^\text{post}$ exhibit level repulsion, indicating that the model is non-integrable \cite{Tur-18a}. Moreover, both initial states have the property that their energy expectation values are $\bra{\psi(0)} \hat{\mathcal{H}}_0^\text{post} \ket{\psi(0)} = 0$, exactly in the middle of the spectrum. Observables, if they thermalise, would therefore be expected to thermalise to their infinite-temperature values in this case. The failure to rapidly thermalise for $\ket{\psi(0)} = \ket{\mathbb{Z}_2}$ was shown to be due to quantum many-body scars, a band of special ETH-violating eigenstates of $\hat{\mathcal{H}}_0^\text{post}$ that have a large overlap with $\ket{\mathbb{Z}_2}$ [see Fig. \ref{fig:1}(c, d), blue circles] \cite{Tur-18a, Tur-18b, Ho-19}.

The large overlap with the special states is reflected in the effective dimension: \begin{equation} D^\text{eff} \equiv \bigg( \sum_{n}|\langle E_n | \psi(0)\rangle|^4 \bigg)^{-1} , \end{equation} where $\ket{E_n}$ are the eigenstates of the post-quench Hamiltonian. Roughly speaking, the effective dimension is the number of distict states through which the system evolves in the course of its dynamics \cite{Lin-09}. We note that this quantity has been used to derive bounds on the fluctuations of observables around their equilibrium values, assuming that the post-quench Hamiltonian has no resonant energy gaps \cite{Pop-06, Rei-08}. We also note that the effective dimension is closely related to the inverse participation ratio (IPR), although the IPR is usually used in the context of localisation of quantum states \cite{Eve-08,Eve-00,Lui-14}. In Fig. \ref{fig:1}(e) we see that $D^\text{eff}$ is much lower for $\ket{\psi(0)} = \ket{\mathbb{Z}_2}$ than for $\ket{\psi(0)} = \ket{\downarrow\downarrow\downarrow\hdots}$, as a result of the large overlap with the quantum many-body scars. Moreover, for $\ket{\psi(0)} = \ket{\downarrow\downarrow\downarrow\hdots}$ the effective dimension increases much more quickly with $L$ than for $\ket{\psi(0)} = \ket{\mathbb{Z}_2}$ (black line vs. blue line). Our goal in this paper is to find physically reasonable initial states that further reduce the effective dimension.




\section{Modifying the pre-quench Hamiltonian}
\label{sec:pre_quench_H}

The initial states $\ket{\psi(0)} = \ket{\mathbb{Z}_2}$ and $\ket{\psi(0)} = \ket{\downarrow\downarrow\hdots}$ considered in the previous section can be represented as ground states of the pre-quench Hamiltonian: \begin{equation} \hat{\mathcal{H}}_0^\text{pre} = \hat{P} \hat{h}_0^\text{pre} \hat{P} , \quad \hat{h}_0^\text{pre} = \frac{\hbar\Delta}{2} \sum_{j=1}^L \hat{\sigma}_j^z . \label{eq:H_0_pre} \end{equation} Choosing $\Delta > 0$ gives the ground state $\ket{\downarrow\downarrow\downarrow\hdots}$. Without the projector $\hat{P}$ implementing the Rydberg blockade the ground state for $\Delta < 0$ would be $\ket{\uparrow\uparrow\uparrow\hdots}$. However, with the blockade, consecutive $\uparrow$-states are forbidden, and choosing $\Delta < 0$ gives the degenerate ground states $(\ket{\mathbb{Z}_2} \pm \ket{\mathbb{Z}'_2})/\sqrt{2}$, with the ``+'' state $(\ket{\mathbb{Z}_2} + \ket{\mathbb{Z}'_2})/\sqrt{2}$ in the $(k,p)=(0,1)$ symmetry sector and the ``-'' state $(\ket{\mathbb{Z}_2} - \ket{\mathbb{Z}'_2})/\sqrt{2}$ in the $(k,p) = (\pi,-1)$ symmetry sector. In any experiment we expect to see spontaneous breaking of the translation and reflection symmetries, giving one of the two N\'{e}el states $\ket{\mathbb{Z}_2}$ or $\ket{\mathbb{Z}'_2}$ as the ground state. This is discussed in more detail in Appendix \ref{app:gs_props}.




It is natural to ask if one can further reduce the effective dimension $D^\text{eff}$, with a physically plausible deformation $ \hat{\mathcal{H}}_0^\text{pre} \to  \hat{\mathcal{H}}^\text{pre}$ of the pre-quench Hamiltonian. To ensure the the pre-quench Hamiltonian $\hat{\mathcal{H}}^\text{pre}$ is physically reasonable we restrict to \emph{local} deformations of $\hat{\mathcal{H}}_0^\text{pre}$. We also assume that the new pre-quench Hamiltonian preserves both the translation-invariance and reflection-invariance of the original pre-quench Hamiltonian. We note that if the ground state $\ket{\psi(0)}$ is an eigenstate of the excitation number parity operator $\hat{\Pi} \equiv \bigotimes_{j=1}^L\hat{\sigma}_j^z$, and if the post-quench Hamiltonian that has particle-hole symmetry $\{\hat{\mathcal{H}}^\text{post}, \hat{\Pi} \} = 0$, then we have \begin{eqnarray} \bra{\psi(0)}\hat{\mathcal{H}}^\text{post}\ket{\psi(0)} &=& \bra{\psi(0)} \hat{\Pi} \hat{\mathcal{H}}^\text{post} \hat{\Pi} \ket{\psi(0)} \\ &=& - \bra{\psi(0)}\hat{\mathcal{H}}^\text{post}\ket{\psi(0)} \end{eqnarray} and so $\bra{\psi(0)}\hat{\mathcal{H}}^\text{post}\ket{\psi(0)} = 0$. Thus, it is convenient to require the pre-quench Hamiltonian to also have parity symmetry $[\hat{\mathcal{H}}^\text{pre}, \hat{\Pi} ] = 0$, since this pins the energy expectation value of the initial state to the middle of the spectrum of $\hat{\mathcal{H}}^\text{post}$. We search for local deformations $\hat{\mathcal{H}}_0^\text{pre} \to \hat{\mathcal{H}}^\text{pre}$ that satisfy our criteria above, and that have the effect of reducing $D^\text{eff}$.

After numerically testing various nearest neighbour and next-nearest neighbour terms, we find that the deformation of the form $\hat{\mathcal{H}}_0^\text{pre} \to \hat{\mathcal{H}}^\text{pre} = \hat{P} \hat{h}^\text{pre} \hat{P}$, where \begin{eqnarray} \hat{h}^\text{pre} &=& \hat{h}_0^\text{pre} + \delta \hat{h}_2^\text{pre} , \label{eq:H_pre} \\ \delta \hat{h}_2^\text{pre} &=& \hbar\sum_{j=1}^L ( g_2^{x} \hat{\sigma}_j^x \hat{\sigma}_{j+2}^x + g_2^{y} \hat{\sigma}_j^y \hat{\sigma}_{j+2}^y ) , \label{eq:delta_H_pre_2} \end{eqnarray} has the most significant effect in decreasing the effective dimension $D^\text{eff}$. Before showing this, we briefly discuss some of the ground state properties of $\hat{\mathcal{H}}^\text{pre}$ after including the next-nearest neighbour terms, since this ground state will be the initial state for the subsequent dynamics. 

\begin{figure}
\includegraphics[width=\columnwidth]{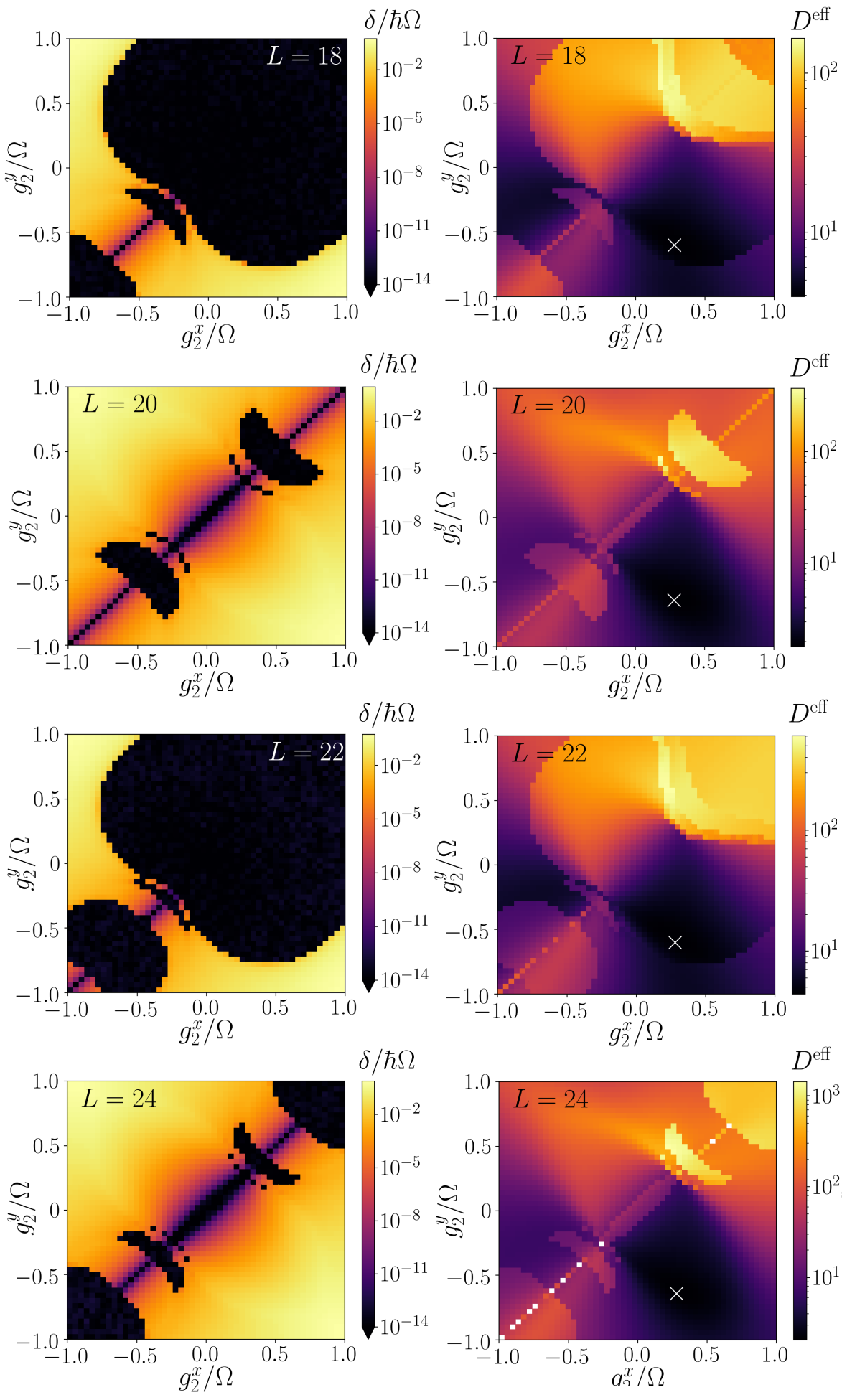}
\caption{Left column: the energy gap $\delta$ between the two lowest energy eigenstates of $\hat{\mathcal{H}}^\text{pre}$ for the Hamiltonian parameters $(g_2^x, g_2^y, \Delta = -1)$ (see Eqs. \ref{eq:H_pre}, \ref{eq:delta_H_pre_2}). Right column: the effective dimension $D^\text{eff}$ for the initial state corresponding to $(g_2^x, g_2^y, \Delta = -1)$, and for the post-quench Hamiltonian $\hat{\mathcal{H}}_0^\text{post}$. The minimum value of $D^\text{eff}$ is marked with a cross. We note, for $L \in \{18,22\}$, discontinuities in $D^\text{eff}$ in the gapless regions are associated with quantum phase transitions of $\hat{\mathcal{H}}^\text{pre}$. However, a further discussion of the quantum phase transition is beyond the scope of this work.}
\label{fig:2}
\end{figure}

A vanishing energy gap $\delta = 0$ between the two lowest eigenstates of $\hat{\mathcal{H}}^\text{pre}$ implies an ambiguity about which superposition of the degenerate states represents the physical ground state. As mentioned previously, this situation already arises for the unperturbed pre-quench Hamiltonian $\hat{\mathcal{H}}_0^\text{pre}$. However, the ambiguity is resolved by spontaneous symmetry breaking, which leads to one of the N\'{e}el states. In Fig. \ref{fig:2} (left column) we plot the energy gap $\delta$ between the two lowest eigenstates of $\hat{\mathcal{H}}^\text{pre}$, for the parameters $(g_2^x, g_2^y, \Delta = -1)$ and for system sizes $L \in \{ 18, 20, 22, 24 \}$. We see that, particularly for $L/2$ odd, there are large regions of parameters that (up to numerical precision) result in a ground state degeneracy $\delta = 0$. In Appendix \ref{app:gs_props} we outline our procedure, based on spontaneous symmetry breaking, for choosing the ground state in our numerical calculations when $\delta = 0$.

\section{Decreased $D^\text{eff}$ and enhanced revivals via modified initial states}
\label{sec:results}

We now show numerically that the modification of the pre-quench Hamiltonian by the next-nearest neighbour deformation Eq. \ref{eq:delta_H_pre_2} can lead to a reduced effective dimension for the dynamics. In Fig. \ref{fig:2} (right column) we plot the effective dimension for initial states corresponding to the pre-quench Hamiltonian parameters $(g_2^x, g_2^y, \Delta = -1)$ for $L \in \{18, 20, 22, 24 \}$. We see that there is a broad region of parameters that result in a decreased $D^\text{eff}$ compared to the unmodified pre-quench Hamiltonian parameters $(g_2^x, g_2^y, \Delta) = (0,0,-1)$. The optimal parameters $(g_2^{x,\text{opt}}, g_2^{y,\text{opt}},\Delta=-1)$ that minimimise the effective dimension are marked with a cross. By refering to the left column of Fig. \ref{fig:2}, we see that if $L/2$ is odd, the optimal parameters $(g_2^{x,\text{opt}}, g_2^{y,\text{opt}},\Delta=-1)$ fall within the region of ground state degeneracy, but that if $L/2$ is even the optimal ground state is in a non-degenerate phase. 

Not only is $D^\text{eff}$ at its minimum value for $(g_2^{x,\text{opt}}, g_2^{y,\text{opt}},\Delta=-1)$, but Figs. \ref{fig:1}(a, b) show that it also leads to a significant enhancement of the Loschmidt revivals, and a slowdown in their decay (red lines vs. blue lines). There is a qualitative difference in the dynamics depending on whether $L/2$ is odd or even [Fig. \ref{fig:1}(a) vs. Fig. \ref{fig:1}(b)], due to the initial state belonging to different ground state phases of $\hat{\mathcal{H}}^\text{pre}$ in either case. If $L/2$ is odd [Fig. \ref{fig:1}(b)], the Loschmidt probability decays to zero, but revives periodically. On the other hand, if $L/2$ is even, the Loschmidt probability fails to decay completely to zero, and has a revival frequency that is double that of $L/2$ odd.

To understand these features of the dynamics we plot the overlaps of the initial state at the optimal point with the eigenstates of the post-quench Hamiltonian $\hat{\mathcal{H}}_0^\text{post}$. Figs. \ref{fig:1}(c, d) show shows that our modification of the pre-quench Hamiltonian has the effect of increasing the overlap of the initial state with the scar states nearest to the middle of the spectrum, while decreasing the overlap with the scar states further from the middle of the spectrum [red triangles vs. blue circles]. For $L/2$ even, the overlap is dominated by a single scar state at zero-energy, with all other overlaps at least an order of magnitude smaller [Fig. \ref{fig:1}(c)]. The dynamics is therefore partly ``stuck'' in this eigenstate and cannot completely evolve to an orthogonal state, so that $\mathcal{L}(t)$ cannot decay completely. If $L/2$ is odd there are three scar states in the middle of the spectrum that have significant overlap with the initial state [Fig. \ref{fig:1}(d)]. This is sufficient for the initial state to evolve to an approximately orthogonal state, leading to a full decay of $\mathcal{L}(t)$ between its revivals. 

We note that, although our minimisation of the effective dimension leads to significantly enhanced Loschmidt revivals, this does not imply that the revivals are maximised by our perturbation of the initial state. Rather, minimisation of the effective dimension and the maximisation of revivals are two different ways of targeting non-equilibrating initial states. For example, in Fig. \ref{fig:1}(a) the failure of the Loschmidt echo to decay to zero is a feature that emerges directly from the minimisation of the effective dimension and would not have been observed if the focus was solely on maximising the revivals.

We also note that the Loschmidt echo is not likely to be easy to measure experimentally. However, in Appendix \ref{app:local_obs} we show that our perturbation of the initial state also leads to longer-lived oscillations in local observables.




The growth of the effective dimension with the system size $L$ is shown in Fig. \ref{fig:1}(e). For the thermalising initial state $\ket{\psi(0)} = \ket{\downarrow\downarrow\downarrow\hdots}$ [$(g_2^x, g_2^y, \Delta) = (0,0,+1)$], we see that the effective dimension increases exponentially with $L$. For the non-thermalising N\'{e}el state [$(g_2^x, g_2^y, \Delta) = (0,0,-1)$] the rate of increase is much slower, though it appears to be slightly faster than linear in $L$, since the plot of $D^\text{eff}/L$ is slightly increasing in $L$. For the optimal initial state [$(g_2^{x,\text{opt}}, g_2^{y,\text{opt}}, \Delta = -1)$], the increase of the effective dimension is slower still, and appears to be sub-linear in $L$ for the system sizes accessible with our numerics. Despite being a quantum chaotic system, the effective dimension is just $D^\text{eff} \approx 2.3$ for the optimal initial state at $L = 28$. For comparison, the $(k,p) = (0,1)$ symmetry sector in which the dynamics takes place has a much larger dimension $D = 13,201$ [excluding the trivial zero-energy eigenstates $\hat{P}\ket{E} = 0$], showing that the dynamics takes place in a very small fraction of the accessible Hilbert space. 

\section{Conclusion}

In this paper we have shown that the slowdown of equilibration due to quantum many-body scars can be enhanced by local deformations of the pre-quench Hamiltonian (i.e., the initial state). This complements recent results by Choi \emph{et. al.} which have shown that (for the initial N\'{e}el state) revivals can also be enhanced by local deformations of the \emph{post-quench} Hamiltonian \cite{Cho-19}. Our approach to identifying slowly equilibrating initial states is to minimise the effective dimension $D^\text{eff}$. We note that a different approach, based on a time-dependent variational principle, was also recently developed in Ref. \cite{Mic-20}.

  We have found that the most significant decrease in $D^\text{eff}$ is achieved by adding next-nearest neigbour interactions to the pre-quench Hamiltonian. The absence of nearest neighbour terms here is unusual from the point of view of experimental implementation, since interaction strength usually does not increase with increasing distance. However, our pre-quench Hamiltonian with next-nearest neighbour interactions on the spin-1/2 chain can be mapped exactly onto a spin-1 chain with \emph{nearest neighbour} interactions, by blocking neigbouring pairs of spin-1/2 particles together as a single spin-1 (the $\ket{\uparrow\uparrow}$ basis state is excluded by the Rydberg blockade) \cite{Lin-19}. The spin-1 model may therefore be more relevent for experimental implementation of the pre-quench Hamiltonian.

Possible avenues of future research include expanding the range of the interaction in the pre-quench Hamiltonian, and modifying the post-quench Hamiltonian \cite{Cho-19} and the pre-quench Hamiltonian together.

\begin{acknowledgments}
The authors acknowledge support from Science Foundation Ireland through Career Development Award 15/CDA/3240. Computational facilities were provided by the DJEI/DES/SFI/HEA Irish Centre for High-End Computing (ICHEC) through Class C Project dsphy013c. The authors also wish to thank Masud Haque for helpful discussions.
\end{acknowledgments}


\bibliography{/Users/dooleysh/Google_Drive/physics/papers/bibtex_library/refs}


\onecolumngrid
\appendix

\section{Ground state properties of the pre-quench Hamiltonian}
\label{app:gs_props}




For a pre-quench Hamiltonian $\hat{\mathcal{H}}^\text{pre}$ with vanishing energy gap $\delta = 0$, there is an ambiguity about which superposition of the degenerate states represents the true ground state. Our numerics show that for the $\hat{\mathcal{H}}^\text{pre}$ specified by Eqs. \ref{eq:H_pre}, \ref{eq:delta_H_pre_2} one of the two degenerate ground states $\ket{\psi_{p=1}}$ always belongs to the even-parity ($p = 1$) symmetry sector, while the other degenerate ground state $\ket{\psi_{p=-1}}$ belongs to the odd-parity ($p = -1$) symmetry sector. To see the role of spontaneous symmetry breaking, we consider the splitting of the degeneracy by the local perturbation $\hat{\mathcal{H}}^\text{pre} \to \hat{\mathcal{H}}^\text{pre} + \epsilon \hat{\mathcal{H}}^\text{pert}$, where $\hat{\mathcal{H}}^\text{pert} =  \hbar\sum_{j=1}^L (-1)^j \hat{P}\hat{\sigma}_j^z \hat{P}$ is a ``staggered potential'', alternating in sign with the chain site index, and $\epsilon$ is arbitrarily small. By degenerate perturbation theory, the first-order perturbed eigenstates are found by diagonalising the staggered perturbation in the degenerate subspace, i.e., by diagonalising the $2 \times 2$ Hermitian matrix \begin{equation} \left(\begin{array}{cc} \epsilon \bra{\psi_{p=1}}\hat{\mathcal{H}}^\text{pert}\ket{\psi_{p=1}} & \epsilon \bra{\psi_{p=1}}\hat{\mathcal{H}}^\text{pert}\ket{\psi_{p=-1}} \\ \epsilon \bra{\psi_{p=-1}}\hat{\mathcal{H}}^\text{pert}\ket{\psi_{p=1}} & \epsilon \bra{\psi_{p=-1}}\hat{\mathcal{H}}^\text{pert}\ket{\psi_{p=-1}} \end{array}\right) . \label{eq:deg_pert} \end{equation} However, we use the fact that $\ket{\psi_{p=\pm 1}}$ are eigenstates of the spatial reflection operator $\hat{R}$, as well as the identity $\hat{R}^\dagger \hat{\mathcal{H}}^\text{pert} \hat{R} = - (-1)^L \hat{\mathcal{H}}^\text{pert}$, to show that the diagonal matrix elements vanish when $L$ is even: \begin{eqnarray} \bra{\psi_{p=\pm 1}}\hat{\mathcal{H}}^\text{pert}\ket{\psi_{p=\pm 1}} &=& \bra{\psi_{p=\pm 1}}\hat{R}^\dagger \hat{\mathcal{H}}^\text{pert} \hat{R} \ket{\psi_{p=\pm 1}} \nonumber\\ &=& - \bra{\psi_{p=\pm 1}}\hat{\mathcal{H}}^\text{pert}\ket{\psi_{p=\pm 1}} \nonumber\\ &=& 0 . \nonumber \end{eqnarray} Since, for the Hamiltonian $\hat{\mathcal{H}}^\text{pre}$, the eigenstates $\ket{\psi_{p=\pm 1}}$ can always be chosen to be real \cite{San-10}, the off-diagonal elements of the $2 \times 2$ matrix are real, and equal to each other $x \equiv \bra{\psi_{p=1}}\hat{\mathcal{H}}^\text{pert}\ket{\psi_{p=-1}} = \bra{\psi_{p=-1}}\hat{\mathcal{H}}^\text{pert}\ket{\psi_{p=1}}$. Thus, the matrix Eq. \ref{eq:deg_pert} is equal to: \begin{equation} \left(\begin{array}{cc} 0 & \epsilon x \\ \epsilon x & 0 \end{array}\right) , \end{equation} and has the energy eigenstates: \begin{equation} \ket{\psi_\pm} = \frac{1}{\sqrt{2}} \left( \ket{\psi_{p=1}} \pm \frac{\epsilon x}{|\epsilon x|} \ket{\psi_{p=-1}} \right) . \end{equation} Which of the two states $\ket{\psi_\pm}$ has the lower energy depends on the sign of $\epsilon x$. In this paper, if there is a ground state degeneracy we assume that $\epsilon x \to 0^{-}$ tends to zero from below, and that the ground state is $\ket{\psi(0)} = \ket{\psi_+}$. By this procedure we calculate the appropriate ground state $\ket{\psi(0)}$ to use in the case of a ground state degeneracy.

\section{Oscillations of local observables}
\label{app:local_obs}

If the ground states of our perturbed pre-quench Hamiltonian are highly entangled states, it is possible that revivals in the initial state might be difficult to see in local observables. Then the enhanced revivals in the Loschmidt echo might not translate into an experimentally observable effect. However, since $\hat{\mathcal{H}}^\text{pre}$ is a local, gapped, 1D Hamiltonian we know that its ground state entanglement entropy obeys an area law \cite{Has-07}. We thus expect that entanglement in the ground state extends across only a few neighbouring sites in the chain, and that there should be local observables that clearly show the revivals of the initial state.


\begin{figure}
  \centering
  \includegraphics[width=\columnwidth]{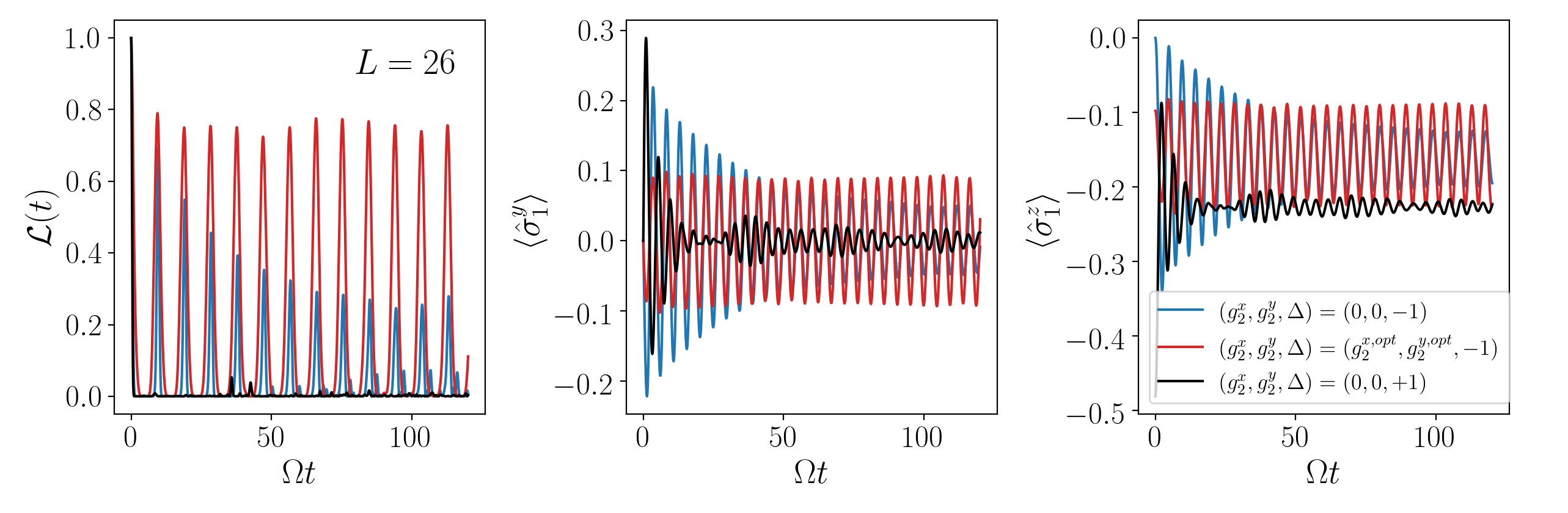}%
\caption{The observables $\langle \hat{\sigma}_1^y \rangle$ and $\langle \hat{\sigma}_1^z \rangle$ undergo long-lived oscillations due to quantum many-body scars, with the lifetime significantly enhanced by our perturbation to the initial state. \label{fig:obs_revivals}}
\end{figure}

In Fig. \ref{fig:obs_revivals} we show the dynamics of the local observables $\hat{\sigma}_1^y$ and $\hat{\sigma}_1^z$. As expected, these observables show long-lived oscillations due to quantum many-body scars, with the lifetime of the oscillations enhanced by our perturbation of the initial state. For short times, the amplitude of the oscillations is reduced compared to the unperturbed initial state $\ket{\mathbb{Z}_2}$. This is probably due to entanglement in the perturbed initial state. However, it may be possible to see larger amplitude oscillation by extending to two-site or three-site local operators.

\end{document}